\documentclass[aps,prl,reprint,showpacs,superscriptaddress]{revtex4-1}
\usepackage{graphicx}
\begin{document}

\title{ 
Theory of fractional-L\'evy kinetics for cold atoms diffusing in optical lattices
 }

\author{D. A. Kessler}
\author{E. Barkai}
\affiliation{Department of Physics, Institute of Nanotechnology and Advanced Materials, Bar Ilan University, Ramat-Gan
52900, Israel}

\begin{abstract}
 Recently,  anomalous  superdiffusion of  ultra cold  $^{87}$Rb atoms
in an optical lattice has been observed along with a fat-tailed, L\'evy type, spatial distribution.
The anomalous exponents were found to depend on the depth of the optical potential. We find, within the framework of
 the semiclassical theory
of Sisyphus cooling,
three distinct phases of the dynamics, as the  optical potential depth is lowered: normal diffusion; L\'evy diffusion; and
$x\sim t^{3/2}$ scaling, the latter related to Obukhov's model
$(1959)$  of turbulence.
 The process can be formulated as a L\'evy walk,
with strong correlations between the length
and duration of the excursions.
We derive a fractional diffusion equation
describing the atomic cloud,
and  the corresponding  anomalous diffusion coefficient.
\end{abstract}
\maketitle
 Very recently, Sagi et al \cite{Sagi} studied  experimentally the diffusion
of ultra-cold $^{87}$Rb atoms in a one dimensional optical lattice.
Starting with a very narrow atomic cloud they recorded the time
evolution of the density
of the particles, here denoted $P(x,t)$ (normalized to unity).
Their work employed the well-known Sisyphus cooling scheme \cite{Dalibard}.
As predicted theoretically  by Marksteiner, et al. \cite{Zoller}, 
the diffusion of the atoms was not Gaussian, so that the
assumption  
that the diffusion process obeys the 
standard  central limit theorem is not valid in this
case. An open challenge is to determine  the precise
nature of the non-equilibrium spreading of the atoms, in particular the dynamical phase diagram
of the various different types of behaviors exhibited as the depth of the optical potential is varied.
In  \cite{Sagi}   
the anomalous diffusion data was compared to the set of 
solutions of the fractional diffusion equation \cite{Saichev,MBK,Review}
\begin{equation}
{\partial^\beta  P(x,t) \over \partial t^\beta} = K_\nu \nabla^\nu P(x,t),  
\label{eq01} 
\end{equation}
with $\beta = 1$, 
so that the time derivative
on the left hand side is a first-order derivative. The fractional
space derivative on the right hand side is a Weyl-Rietz fractional
derivative~\cite{Review}. Here
the anomalous diffusion coefficient 
 $K_\nu$ has units $\mbox{cm}^\nu/sec$. A fundamental challenge is to derive fractional equations from
 a microscopic theory, without invoking power-law statistics in the first place.  Furthermore,
 the solutions of such equations exhibits diverging mean-square displacement $\langle x^2 \rangle = \infty$,
 which violates the principle of causality~\footnote{See the discussion in Ref. \cite{KSZ} where the unphysical nature of L\'evy flights is discussed and the resolution in terms of L\'evy walks is addressed.}, which permits physical phenomena to spread at finite
 speeds.  So how can fractional equations like Eq. (\ref{eq01}) describe physical reality?  We will address this paradox in this work.
 
The solution of Eq. 
(\ref{eq01}) for an initial narrow cloud 
is given in terms of a L\'evy distributions (see details below).
The L\'evy distribution generalizes the Gaussian distribution in the
mathematical problem of the sum of a large number of 
independent random variables, in the case where the variance of 
summands diverges, physically corresponding to 
scale free systems.
 L\'evy statistics
and fractional kinetic equations  have found several applications 
\cite{Review,Bouchaud,SZK,KSZ,Stanley,BNVBK,MPKB,Barthelemy,Humphries},
including in the context of sub-recoil laser cooling \cite{Levybook}. Here
our aim is to derive  L\'evy statistics and the fractional diffusion equation
from the semi-classical picture of Sisyphus cooling. Specifically we will show that $\beta=1$
and relate the value of the exponent $\nu$ to the depth
of the optical lattice $U_0$, deriving an expression for the
constant $K_\nu$. Furthermore, we discuss the limitations of the
fractional framework, and show that for a critical value of
the depth of the optical lattice,
the dynamics switches to a non-L\'evy behavior (i.e. a regime where
Eq. (\ref{eq01}) is not valid); instead it is related to Richardson-Obukhov diffusion found in turbulence.  
Thus the semiclassical picture predicts
a rich phase diagram for  the
atomistic
diffusion process. We will then compare the results of this analysis to the experimental
findings, and see that there are still unresolved discrepancies between the experiment and the theory.  Reconciling the two thus poses
a major challenge for the future.

{\em Model and goal.} In this article we investigate the spatial
 density of the atoms,
$P(x,t)$.
The trajectory of a single particle is $x(t) = \int_0 ^t p(t) {\rm d} t/m$ where
$p(t)$ is its momentum.
Within the standard semiclassical picture \cite{Dalibard,Zoller}
of Sisyphus cooling, two competing
mechanisms describe the dynamics. 
The cooling force 
$F(p) = - \overline{\alpha} p /[1 + (p/p_c)^2 ]$
acts to restore the momentum to the minimum energy state $p=0$.
Momentum diffusion is governed by  a
diffusion coefficient which is momentum dependent,
$D(p) = D_1 + D_2 / [1 + (p/p_c)^2]$. The latter
describes momentum fluctuations which
lead to heating (due to random emission events). 
We use dimensionless units, time $t \to
t \overline{\alpha}$, momentum $p \to p/p_c$,    the momentum diffusion
constant  $D=D_1/ (p_c)^2 \overline{\alpha}$ and 
$x \to x m \overline{\alpha}  / p_c$. 
For simplicity,
we set $D_2=0$  since
 it does not modify the asymptotic $|p|\to \infty$
 behavior of the
diffusive heating term, nor that of the force
 and therefore  does not modify our main conclusions.
The Langevin equations
\begin{equation}
{{\rm d} p \over {\rm d} t } = F(p) + \sqrt{ 2 D} \xi(t), \ \ \ \ \ \ 
{{\rm d} x \over {\rm d} t } = p  
\label{eq05} 
\end{equation} 
describe the dynamics in phase space. 
Here the noise term is Gaussian, has zero mean
and is white $\langle \xi(t) \xi(t') \rangle = \delta(t- t')$. 
The now dimensionless cooling force is
\begin{equation}
F(p) = - { p \over 1 + p^2} . 
\label{eq03}
\end{equation}
The stochastic Eq. (\ref{eq05})
gives the trajectories of the standard  
Kramers picture  for   the semi-classical dynamics in the optical lattice
which in turn was derived from microscopical considerations \cite{Dalibard,Zoller}. 
 From the semiclassical treatment of the interaction of the atoms with
the counter-propagating laser beams, we have 
$D= c E_R/ U_0$, 
where $U_0$ is the depth of the optical potential and  $E_R$
the recoil energy, and the dimensionless parameter
$c$~\footnote{For atoms in molasses 
with a $J_g=1/2 \rightarrow J_e =3/2$
Zeeman substructure in a lin $\bot$ lin laser configuration $c=12.3$
\cite{Zoller}.
Refs. \cite{Dalibard,Lutz,Renzoni}
give  $c=22$. 
As pointed out  in \cite{Zoller} different notations are used in
the literature.}
depends on the atomic transition involved \cite{Dalibard,Zoller,Lutz}.
 For $p \ll 1$, the cooling force Eq. (\ref{eq03})  
is  harmonic, $F(p) \sim - p$, 
while in the opposite limit, $p\gg 1$, $F(p) \sim - 1/p$. The effective
potential $V(p) = - \int_0 ^p F(p) {\rm d} p= (1/2) \ln(1 + p^2)$
is asymptotically logarithmic, $V(p) \sim \ln(p)$
when $p$ is large. This large $p$ behavior of $V(p)$
is responsible
for several unusual equilibrium and non-equilibrium
properties of the momentum distribution
\cite{Katori,Renzoni,KesslerPRL,Mukamel2} 
while the new experiment \cite{Sagi}
demands a theory
for the spatial spreading.

\begin{figure}\begin{center}
\includegraphics[width=0.5\textwidth]{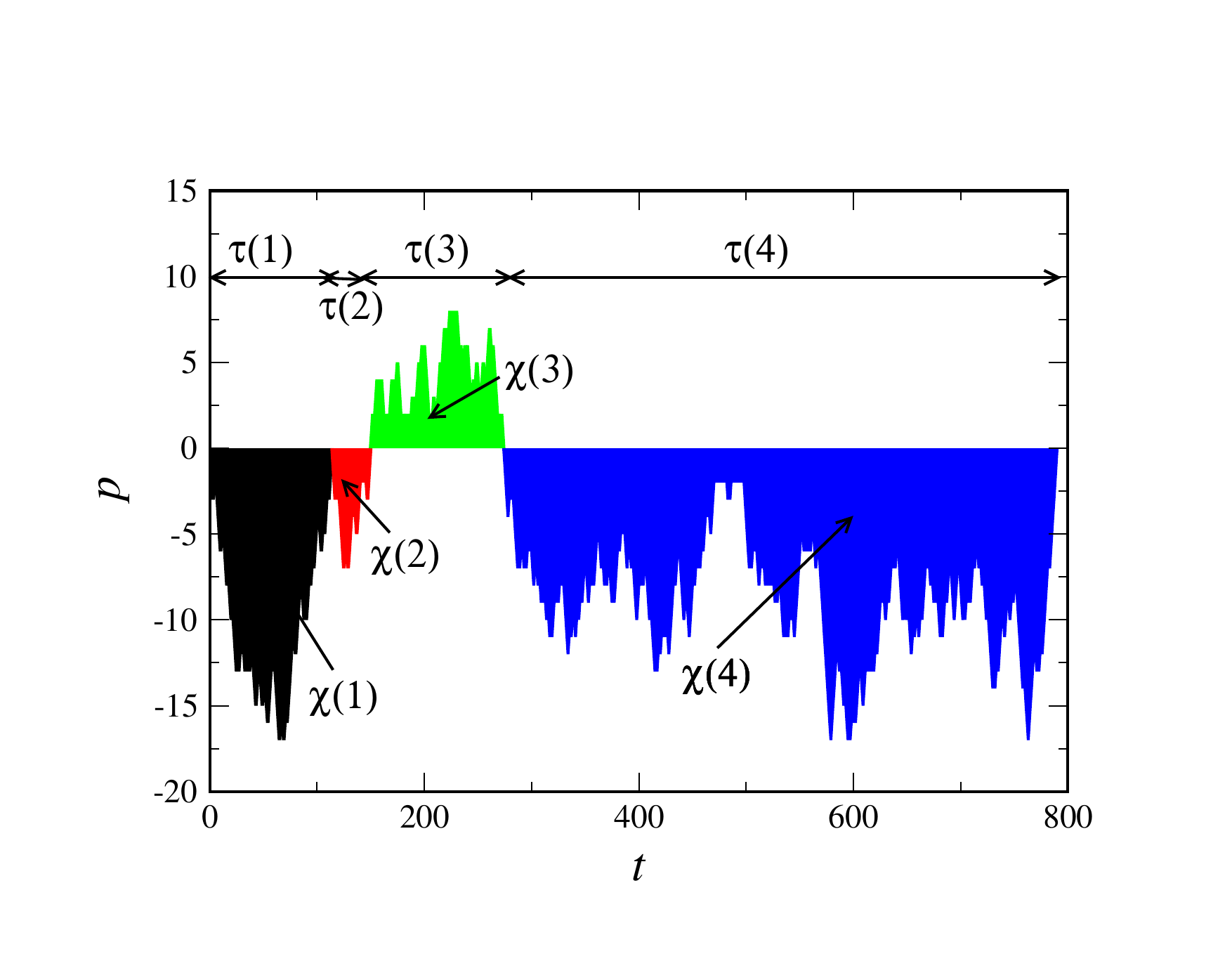}
\end{center}
\caption{
Schematic presentation of momentum of the particle versus time.
The times between consecutive
zero crossings are called the jump durations $\tau$ and the shaded
area under each excursion are the random flight displacements $\chi$.
The $\tau$'s and the $\chi$'s are correlated, since statistically
a long jump duration implies a large displacement.
}
\label{fig1}
\end{figure}

 The heart of our analysis is the mapping of the Langevin dynamics to a recurrent set of random walks.
 The particle along its stochastic path in momentum space
crosses $p=0$ many times when the measurement time is long.
Let $\tau>0$ be the random time between one crossing event to
the next crossing event,
and let $-\infty < {\chi}  < \infty$ be the
random displacement (for the corresponding $\tau$).
 As schematically shown in Fig. \ref{fig1},
the process starting at the origin with
zero momentum is defined by the sequence of jump durations, $\left\{ \tau(1), \tau(2) , ....\right\}$ with corresponding
displacements $\left\{ {\chi(1)}, {\chi(2)}, \cdots \right\}$, with
$\chi(1) \equiv \int_0^{\tau(1)} p(\tau) d\tau$, $\chi(2) \equiv \int_{\tau(1)}^{\tau(1)+\tau(2)} p(\tau)d\tau$, etc.
 The total 
displacement $x$ at time $t$ is a sum of the individual
displacements ${\chi(i)}$. 
 Since the underlying Langevin process
is continuous, 
we need a more precise definition of this process. 
We define $\tau$ as  the time it takes the particle
with initial momentum $p_i$ to reach $p_f=0$ for the 
first time, where eventually we take  $p_i \to p_f$.
Similarly, ${\chi}$ is the displacement of the particle 
during this flight. 
The probability density function (PDF) of the displacement
$\chi$ is denoted $q({\chi})$ and
of the jump durations $g(\tau)$. 

As shown by Marksteiner, et al. \cite{Zoller}, these PDFs 
exhibit power law behavior
\begin{equation}
g\left( \tau \right) \propto \tau^{-{ 3\over 2} - {1\over 2 D}},\ \ \  q\left({\chi}\right) \propto |{\chi}|^{ - {4 \over 3}  - {1 \over 3 D} }, 
\label{eq06}
\end{equation}
as a consequence of the logarithmic potential,
which makes the diffusion for large enough $p$ only weakly bounded.  It is this power-law behavior, with its divergent second moment of the displacement $\chi$ for $D>1/5$, which gives rise to
the anomalous statistics for $x$.  
Importantly, and previously overlooked, there is a 
strong correlation between the jump duration $\tau$ and
the spatial extent of the jumps $\chi$. 
These correlations have important consequences, including the
finiteness of the moments of $P(x,t)$ and the $D>1$ dynamical phase we 
obtain below.  These correlations are responsible,
in particular, for the finiteness of the moments of $P(x,t)$.
Physically, such a correlation is obvious, since long
jump durations involve large momenta, which in turn
induce a large spatial displacement.
 The theoretical development starts then from the quantity 
 $\psi({\chi}, \tau)$,
the joint probability density of  ${\chi}$ 
and $\tau$. 
From this, we construct a L\'evy walk
scheme \cite{KBS,Blumen,Carry} which relates the microscopic
information $\psi(\chi,\tau)$ to the atomic packet $P(x,t)$ for large $x$ and $t$.

\begin{figure}\begin{center}
\includegraphics[width=0.5\textwidth]{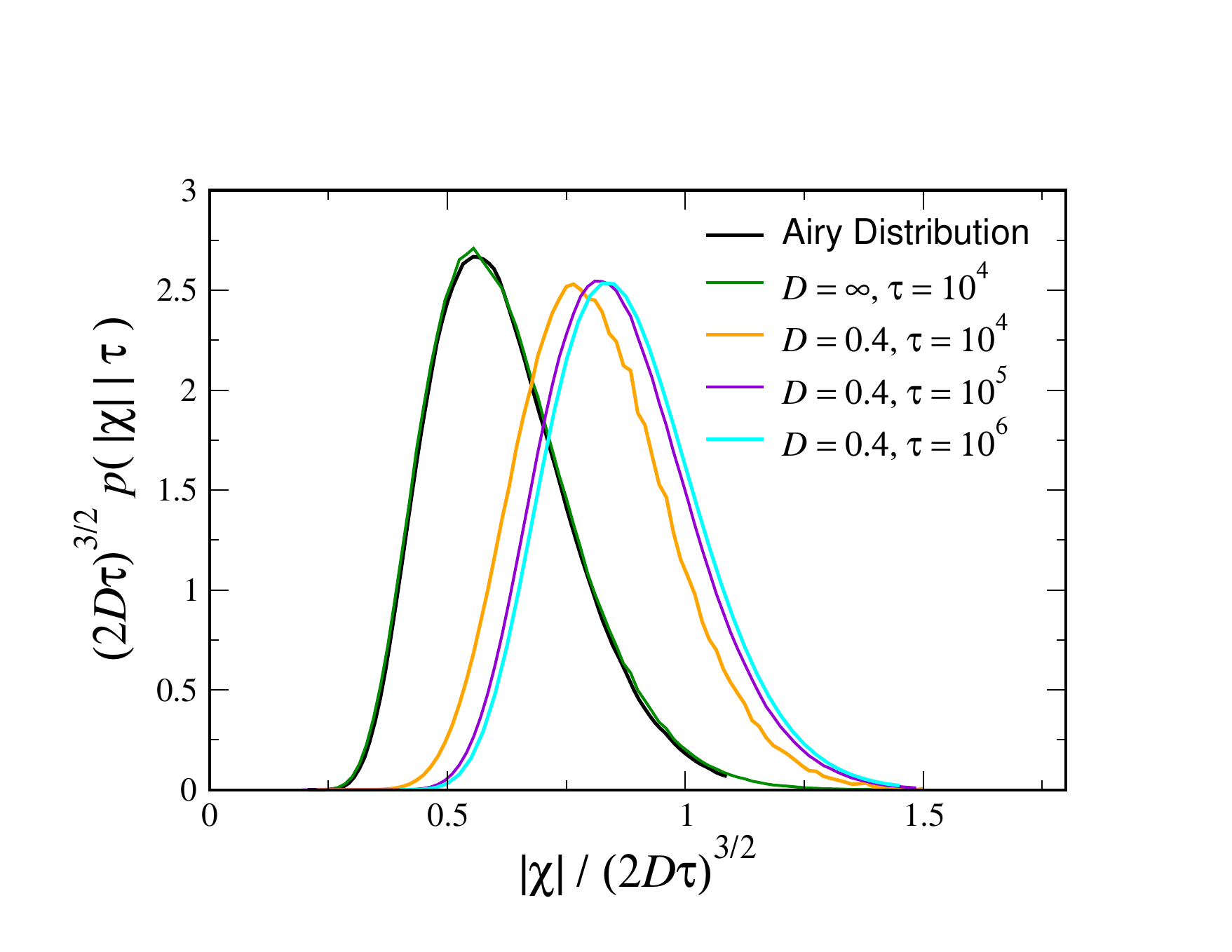}
\end{center}
\caption{
The scaled conditional probability $(2D\tau)^{3/2} p(|\chi|\,\big{|}\,\tau)$ versus $|\chi|/(2D\tau)^{3/2}$  for $\tau = 10^4$, $10^5$, and $10^6$,
for the case $D=0.4$, from simulations, showing the convergence to an asymptotic scaling form.  Also shown is the $D\to \infty$  limit for $\tau=10^4$, as well as the analytic result for $\tau\to\infty$, the Airy distribution~\cite{Majumdar,Majumdar2}.
}
\label{fig2}
\end{figure}

{\em Scaling theory for anomalous diffusion.} We rewrite the joint PDF 
$\psi( \chi, \tau)=g(\tau) p( \chi |\tau)$,
where $p(\chi \, | \,\tau)$ is the conditional probability to find a jump length of 
$ \chi$ for a given jump duration $\tau$. 
Numerically, as shown if Fig. \ref{fig2}, we observed that the conditional probability scales
at large times like 
\begin{equation}
p(\chi |\tau) \sim \tau^{- \gamma} B(\chi / \tau^\gamma)
\label{eqSca}
\end{equation}
and $\gamma=3/2$ and $B(\cdot)$ is a scaling function.
 To analytically obtain  the scaling exponent $\gamma=3/2$ note
that
$q\left({\chi}\right)=\int_0 ^\infty {\rm d} \tau \psi \left( \chi,\tau \right)$, giving%
\begin{equation} 
q \left( {\chi} \right) \sim 
 \int_{\tau_0}  ^\infty {\rm d} \tau \tau^{-{3\over 2 }- {1 \over 2 D}}  \tau^{- \gamma} B \left(\frac{ \chi  }{ \tau^\gamma}\right) 
 \propto |{\chi}|^{ - \left(1 + \frac{1+1/D}{2\gamma}\right)} .
\label{eq07} 
\end{equation}
Here $\tau_0$  is a time scale after which the long time
limit in Eq. (\ref{eq06}) holds and is irrelevant for large $\chi$.
Comparing Eq. (\ref{eq07}) to the second of  Eqs. 
(\ref{eq06})  yields 
   the consistency condition
 $1 + (1 + 1/D)/(2 \gamma) =
4/3 + 1/ (3 D)$ and hence $\gamma=3/2$, as we observe in Fig. \ref{fig2}.

It is interesting to note that $p(\chi|\tau)$  in the case of free diffusion (corresponding to the limit $D\to\infty$) has been previously considered by mathematicians~\cite{Darling,Louchard,Takacs} and shown to obey the
scaling relation Eq. (\ref{eqSca}), with $B$  given by the so-called Airy distribution~\cite{Majumdar,Majumdar2}.  In the case of finite $D$, an analytic formula for $p(\chi|\tau)$ can be constructed using the Feynman-Kac formalism~\cite{future}, giving for asymptotically long walks both $\gamma=3/2$ and a closed form expression for the scaling function $B(\cdot)$.
For our current purposes, however, we do not need the exact form of $B$; what is important is the scaling behavior, and the fact that,
 $B$ falls of rapidly for large argument, ensuring
finite moments of this function. 
 
Given our scaling solution for 
$p(\chi|t)$, and hence $\psi(\chi,t)$, the next step is to 
construct a theory for the spreading 
of the particle packet using tools developed in the random walk community
\cite{KBS,Blumen,Carry}.  One first obtains~\cite{future} a Montroll-Weiss \cite{Review} type of equation for the Fourier-Laplace transform of $P(x,t)$, $\widetilde{P}(k,u)$, in terms of $\widetilde{\psi}(k,u)$,
the Fourier-Laplace transform of the joint PDF $\psi(\chi,\tau)$:
\begin{equation}
\widetilde{P}(k,u) = \frac{\Psi(k,u)}{1-\widetilde{\psi}(k,u)} . 
\label{MW}
\end{equation}
Here, $\Psi(k,u)$ is the Fourier-Laplace transform of $\tau^{-3/2}B(|\chi|/\tau^{3/2})[1-\int_0^t \psi(\tau) d\tau]$. The last step is then to invert Eq. (\ref{MW})
back to the $x$, $t$ domain.

\begin{figure}\begin{center}
\includegraphics[width=0.5\textwidth]{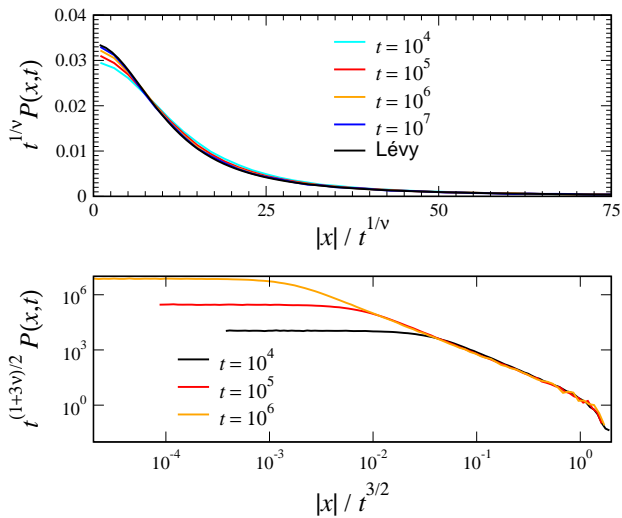}\\
\end{center}
\caption{Upper panel: 
$t^{1/\nu} P(|x|,t)$ versus $|x|/t^{1/\nu}$ for $D=2/5$, i.e. $\nu=7/6$. The theory:
L\'evy PDF Eq. (\ref{eq18}) with $K_\nu$ Eq. (\ref{eq21}), perfectly matches
simulations without fitting.  
Lower panel:  $t^{(1+3\nu)/2} P(|x|,t)$ versus $|x|/t^{3/2}$ for $D=2/5$, showing the universal crossover
from power-law to Gaussian behavior at $|x| \sim t^{3/2}$
}
\label{fig3}
\end{figure}

We now explain why L\'evy statistics describes the diffusion profile $P(x,t)$ when $1/5 < D < 1$, provided that $x$ is not too large.
The key idea is  that, for $x$'s which are large, but not extremely large, the problem decouples, and $\widetilde{\psi}(k,u)$ can be expressed as a product of the
Fourier transform of $q(\chi)$, $\tilde{q}(k)$ and the Laplace transform of $g(\tau)$, $\tilde{g}(u)$.  This is valid as long as $x \ll t^{3/2}$, since otherwise
paths where $\chi \sim t^{3/2}$ are relevant, for which the correlations are strong, as we have seen.  The long-time, large-$x$ behavior of $P(x,t)$ in the
decoupled regime is then governed by
the small-$k$ behavior of  $\tilde{q}(k)$ and the small $u$ behavior of  $\tilde{g}(u)$.    When the second moment of $q(\chi)$ diverges, i.e. for $D>1/5$, the small-$k$ behavior  of $\tilde{q}(k)$ is determined by the large-$\chi$ asymptotics of $q(\chi)$ as given in Eq. (\ref{eqSca}),
$q(\chi) \sim  x_*^\nu/|\chi|^{1+\nu}$, where we have introduced the parameter
\begin{equation}
\nu \equiv \frac{1+D}{3D}\ .  
\end{equation}When the first moment of $\tau$ is finite, i.e. for $D<1$, the small-$u$ behavior of  $\tilde{g}(u)$ is
determined by the first moment, $\langle \tau \rangle$:  $\tilde{g}(u) \sim 1 - u\langle\tau\rangle$.  From these follow the small-$k$, small-$u$ behavior of $\widetilde{P}(k,u)$:
\begin{equation}
\widetilde{P}(k,u) \sim \frac{1}{ u +  K_\nu |k|^\nu}
\label{Levyku}
\end{equation}
where $K_\nu = \pi x_*^\nu / (\langle \tau \rangle \Gamma(1+\nu)\sin \frac{\pi\nu}{2})$.  Both $x_*^\nu$ and $\langle \tau\rangle$ can be calculated via appropriate backward Fokker-Planck equations.  They both vanish as the magnitude of the initial momentum of the walk goes to zero, but their ratio has a finite limit, so that $K_\nu$, upon returning to dimensionfull units, is
\begin{equation}
K_\nu = { \sqrt{\pi} (3 \nu -1)^{\nu -1} \Gamma\left(\frac{3\nu-1}{2}\right) \over \Gamma\left(\frac{3\nu-2}{2}\right) 3^{2 \nu-1} [\Gamma(\nu)]^2 \sin\left( \frac{\pi \nu}{2} \right)}
\left({p_c \over m} \right)^\nu ( \overline{\alpha})^{- \nu + 1}.
\label{eq21}
\end{equation}
$\widetilde{P}(k,u)$, as given in Eq. (\ref{Levyku}), is in fact precisely the symmetric L\'evy distribution in Laplace-Fourier space with index $\nu$, whose $(x,t)$ representation is (see Eq. (B17) of \cite{Bouchaud})
\begin{equation}
P(x,t) \sim {1 \over \left(K_\nu t\right)^{1/\nu}}   L_{\nu,0}\left[{x \over
\left( K_\nu  t^{1/\nu}\right)} \right] 
\label{eq18}
\end{equation}
It is easy to see that this distribution is the solution of 
 the fractional diffusion equation, Eq. (\ref{eq01}), with $\beta=1$ and an initial distribution located at the origin.  This justifies the use of Eq. (\ref{eq01}) in
  Ref.~\cite{Sagi} for $1/5 < D < 1$ and provides $\nu$ and $K_\nu$ in terms of the experimental parameters.  We can verify this behavior in simulations, as shown in the upper panel of Fig. \ref{fig3}, where we
see excellent agreement to our theoretical prediction, Eqs. (\ref{eq18}) and (\ref{eq21}), without any fitting.

The lower panel of Fig. 3  illustrates the  cutoff on the L\'evy distribution, which is found at distances $x\sim t^{3/2}$.  Beyond this length scale, the density falls off rapidly.  This, as noted above, is the result of the correlation between $\chi$ and $\tau$, as there are essentially no walks with a displacement greater than the order of $t^{3/2}$.  This cutoff ensures the finiteness
of the mean square displacement, using the power law
tail of the L\'evy PDF $L_\nu(x) \sim x^{- (1 + \nu)}$
 and the cutoff   we get:
 $\langle x^2 \rangle\simeq \int^{t^{3/2}} t^{-(1/\nu)} (x/t^{1/\nu})^{ - (1 + \nu)} x^2 {\rm d} x\sim t^{4 -3 \nu/2}$, for $2/3 < \nu < 2$, in agreement with
\cite{Dechantprl}.
As noted in the introduction, if we rely on the fractional diffusion equation, Eq. (\ref{eq01}), 
naively, we get $\langle x^2 \rangle=\infty$. 
Thus the fractional equation must be used with care,
realizing its limitations in the statistical description of the moments
of the distribution and its tails.  
When $D<1/5$, the diffusion is normal since the variance of the walk displacement is finite.

{\em The Obukhov-Richardson phase, $D>1$.} When 
 the average jump duration, $\langle \tau \rangle$, diverges,  i.e., for $D>1$,
the dynamics of $P(x,t)$ enters a new phase.  Since the L\'evy index $\nu$ approaches 2/3 as $D$ approaches 1,
$x$ scales like $t^{3/2}$ in the limit.  Due to the correlations, $x$ cannot grow faster than this, so in this regime,
$P(x,t) \sim t^{-3/2} h(x/t^{3/2})$, which clearly describes a  correlated phase.
This scaling is that of free diffusion, namely momentum diffuses
like $p \sim t^{1/2}$ and hence the time  integral over the momentum scales
like $x \sim t^{3/2}$. Indeed, in the absence of the logarithmic potential,
namely in the limit $D\gg1$ Eq. (\ref{eq05}) gives    
\begin{equation}
P\left(x,t\right) \sim \sqrt{ {3 \over 4 \pi D t^3}} \exp\left[ - {3 x^2 \over 4 D t^3} \right].
\label{eq22}
\end{equation}
This limit describes the Obukhov model for a tracer particle path in
turbulence, where 
the velocity follows a simple Brownian motion \cite{Obukhov,Friedrich}.
 These scaling properties are
related to Kolmogorov's theory of 1941
(see Eq. (3) in \cite{Friedrich}) and 
to Richardson's diffusion $\langle x^2 \rangle \sim t^3$ 
\cite{Rich,Dechantprl}.
Eq. (\ref{eq22}) is valid when the optical potential depth is small
since $D\to \infty$ when $U_0 \to 0$. 
This limit should be taken with care,
as the observation time must be made large 
before considering the limit of weak potential.
In the opposite scenario, i.e. $U_0 \to 0$ before $t \to \infty$,
we expect ballistic motion, $|x| \sim t$,
since then the optical lattice has not had time to make itself 
felt \cite{Sagi}. 

{\em The relation to Brownian excursions.}
 Our work shows a surprising connection between the dynamics of cold
atoms  to the problem of the area under a Brownian excursion
~\cite{Darling,Louchard,Takacs,Majumdar,Majumdar2}, constrained random walks that start at the origin and return there for the {\em first} time after $t$ steps. 
This non-trivial problem  has
applications in computer science and graph theory and recently
to  the properties of
fluctuating interfaces
 \cite{Majumdar,Majumdar2}. Here it corresponds to the calculation of $p(\chi|\tau)$ for the case of $D\to\infty$ corresponding to free diffusion, since $\chi$ is the area under the random walk path in a single excursion (see Fig. \ref{fig1}).
The current problem with finite $D$ constitutes an interesting generalization of the problem to logarithmically biased random walks, which has the same scaling exponent but a $D$-dependent distribution, which in analogy with the term Brownian excursion, we entitle a Bessel excursion.  The term Bessel excursion stems from the fact that mathematically speaking,
 diffusion in momentum space in a non-regularized potential $\ln(p)$
corresponds to a process called the Bessel process \cite{Bray,Schehr}.
A Bessel excursion  is the Langevin path $p(t')$ 
over the time interval $0\le t' \le \tau$ such that
the path starts on $p_0 \to 0$ and ends on the
origin, but is {\em constrained} to stay positive (if $p_i>0$) or
negative (if $p_i<0$). In \cite{future} we will provide a detailed account
on these constrained random paths.

  {\em Discussion of the experiment and summary.} 
Our work shows a rich phase diagram of the dynamics, with two transition
points. For deep wells, $D<1/5$,
 the diffusion is Gaussian, while for $1/5<D<1$ 
 we have L\'evy statistics, and for $D>1$ 
 Richardson-Obukhov scaling,  $x \sim t^{3/2}$,  is found.
 We have shown that the correlations between jump durations $\tau$ and displacements $\chi$ are
crucial for the behavior of the tails of the  distribution  of the total displacement $x$ and are responsible
for the finiteness of its second moment.
When $D>1$ the correlations become strong, leading to a breakdown of decoupled
L\'evy diffusion.
So far,  experiments have not detected these transitions,
though \cite{Sagi}  clearly demonstrated that 
the change in optical potential depth,
controls the anomalous exponents in the L\'evy spreading packet. 
 In particular so far the experiment
showed at most ballistic behavior, with the spreading exponent $\delta$, defined by $x \sim t^\delta$, always less than unity. This might be related to our
observation that to go beyond ballistic motion, $\delta>1$, one
must take the measurement time to be very long.
A more serious problem is that, in the experimental fitting of the diffusion 
front
to the L\'evy propagator, 
an additional exponent was introduced
 \cite{Sagi} 
to describe the time dependence of the full width at half 
maximum. In contrast, our semi-classical theory shows that a single
exponent $\nu$ is needed within the L\'evy scaling regime $1/5<D<1$, with the spreading exponent $\delta=1/\nu$. 
This might be related to the cutoff of the tails of L\'evy
PDF which demands that the fitting be performed in the central part of the atomic cloud.
 On the other hand we cannot 
rule out other physical effects not included in the semiclassical model. For example
it would be very interesting to simulate the system with quantum Monte Carlo
simulations, though we note that these are not trivial in the 
$|x| \sim t^{3/2}$ regime since the usual simulation procedure 
introduces a cutoff on
momentum, which may give rise to an artificial ballistic motion.  Thus while there is some tantalizing
points of contact between the theory and experiment, achieving full agreement will require more study.

\acknowledgments
This  work  was supported by the  Israel Science  Foundation. 
We thank A. Dechant, E. Lutz, Y. Sagi and N. Davidson for discussions.

\bibliography{bibfile}

\end{document}